\documentclass[preprint,12pt,table,xcdraw]{elsarticle}

\usepackage[T1]{fontenc}
\usepackage[utf8x]{inputenc}
\usepackage[none]{hyphenat}
\usepackage[margin=2.45cm]{geometry}
\usepackage{framed}
\usepackage[inline]{enumitem}
\usepackage{tabularx}
\usepackage{hyperref}
\usepackage{comment}
\usepackage{booktabs} 
\usepackage{graphicx}
\usepackage{array}
\usepackage{float}
\usepackage{adjustbox}
\usepackage{multirow}
\usepackage[table]{xcolor}

\usepackage{listings}
\newcommand{\TODO}[1]{\textcolor{red}{#1}\GenericError{}{LaTeX Error: TODO: #1}}\newcommand\todo\TODO

\newcommand{\rev}[1]{\textcolor{black}{#1}\GenericWarning{}{LaTeX Warning: REV: #1}}

\newcommand{\X}{$\bullet$}
\newcommand{\stack}[2]{\vtop{\hbox{\strut #1}\hbox{\strut #2}}}

\newcommand{\vertical}[1]{\rotatebox[origin=l]{90}{#1}}

\usepackage{lscape}

\usepackage{mathastext}
\usepackage{hyperref}
\hypersetup{
  pdfcreator = {},
  pdfproducer = {}
}

\begin{document}
\title{A Snowballing Literature Study on Test Amplification}
\author{Benjamin Danglot$^1$,
Oscar Vera-Perez$^1$,
Zhongxing Yu$^3$,\\
Andy Zaidman$^2$, Martin Monperrus$^3$, Benoit Baudry$^3$\\
$^1$~INRIA, $^2$~TU Delft, $^3$~KTH Royal Institute of Technology
}
\maketitle

\textbf{Abstract:} The adoption of agile approaches has put an increased emphasis on testing, resulting in extensive test suites. These suites include a large number of tests, in which developers embed knowledge about meaningful input data and expected properties as oracles. 
This article surveys works that exploit this knowledge to enhance manually written tests with respect to an engineering goal (e.g., improve coverage or refine fault localization). While these works rely on various techniques and address various goals, we believe they form an emerging and coherent field of research, which we coin ``test amplification''.
We devised a first set of papers from DBLP, searching for all papers containing "test" and "amplification" in their title. We reviewed the 70 papers in this set and selected the 4 papers that fit the definition of test amplification. We use them as the seeds for our snowballing study, and systematically followed the citation graph. 
This study is the first that draws a comprehensive picture of the different engineering goals proposed in the literature for test amplification.
We believe that this survey will help researchers and practitioners entering this new field to understand more quickly and more deeply the intuitions, concepts and techniques used for test amplification.


\section{Introduction}

Software testing is the art of evaluating an attribute or capability of a program to determine that it meets its required results \cite{hetzel1988}. 

\rev{With the advent of agile development methodologies, which advocate testing early and often, a growing number of software projects develop and maintain a test suite~\cite{Madeyski2010}}. Those test suites are often large and have been written thanks to a lot of human intelligence and domain knowledge~\cite{azaidmanEMSE2011,DBLP:conf/icst/ZaidmanRDD08}. Developers spend a lot of time in writing the tests~\cite{BellerTSE,beller2015when,beller2015howmuch}, so that those tests exercise interesting cases (including corner cases), \rev{and so that an oracle verifies as much as possible the program behavior~\cite{hilton2018coverageevolution}}.

The wide presence of valuable manually written tests has triggered a new thread of research that consists of leveraging the value of existing manually-written tests to achieve a specific engineering goal. This is what we coin ``test amplification''. We introduce the term \emph{amplification} as an umbrella for the various activities that analyze and operate on existing test suites and that are referred to as augmentation, optimization, enrichment, or refactoring in the literature. 

The goal of this paper is to help this original research thread thrive. This has motivated us to conduct a survey of research literature so that existing research efforts are characterized, can be compared and new research opportunities can be identified. Furthermore, it is our conjecture that with good foundations and maturation, test amplification has the potential to bring software testing to the next level in terms of efficiency and efficacy among practitioners by introducing new automatic processes that improve the manually written tests.


This paper studies the literature on test amplification. The reviewing methodology is based on backward- and forward- snowballing on the citation graph \cite{jalali2012systematic}.
To the best of our knowledge, this review is the first that draws a comprehensive picture of the different engineering techniques and goals proposed in the literature for test amplification. 

We structure our reviewed papers in four main categories, each of them being presented in a dedicated section.
Section \ref{sec:amp_add} presents techniques that synthesize new tests from manually-written tests.
Section \ref{sec:amp_change} focuses on the works that synthesize new tests dedicated to a specific change in the application code (in particular a specific commit).
Section \ref{sec:amp_exec} discusses the less-researched, yet powerful idea of modifying the execution of manually-written tests. 
Section \ref{sec:amp_mod} is about the modification of existing tests to improve a specific property.

To sum up, our contributions are:
\begin{itemize}
	\item The first ever snowballing literature review on test amplification
	\item The classification of the related work into four main categories to help newcomers in the field  (students, industry practitioners)  understand this body of work.
	\item A discussion about the outstanding research challenges of test amplification.
\end{itemize}

\section{Method}
In this section, we present the methodology of our systematic literature review.

\subsection{Definition}
\label{sec:core-definition}
In this review, we use the following definition of test amplification:

\begin{framed}
	\textbf{Definition}: Test amplification consists of exploiting the knowledge \rev{of a large number of test cases}, in which developers embed meaningful input data and expected properties in the form of oracles, in order to enhance these manually written tests with respect to an engineering goal (e.g., improve coverage of changes or increase the accuracy of fault localization).
\end{framed}

\emph{Example:} 
A form of test amplification is the addition of test cases automatically generated from the existing manual test cases to increase the coverage of a test suite over the main source code.

\emph{Relation to related work:} 
Test amplification is complementary, yet, significantly different from most works on test generation.
The key difference is what is given as input to the system. Most test generation tools take as input:
the program under test or a formal specification of the testing property.
\textbf{In contrast, test amplification is defined as taking as primary input test cases written by developers}.

\subsection{Methodology}
\label{sec:methodology-step-by-step}

Literature studies typically rigorously follow a methodology to ensure both completeness and replication. We refer to Cooper's book for a general methodological discussion on literature studies~\cite{cooper1998synthesizing}. Specifically for the field of software engineering, well-known methodologies are systematic literature reviews (SLR)~\cite{kitchenham2004procedures}, systematic mapping studies (SMS)~\cite{petersen2008systematic} and snowballing studies~\cite{wohlin2014guidelines}.
For the specific area of \emph{test amplification}, we found that there is no consensus on the terminology used in literature. This is an obstacle to using the SLR and SMS methodologies, which both heavily rely on searching~\cite{Brereton2007}. As snowballing studies are less subject to suffering from the use of diverse terminologies, we perform our study per Wohlin's guidelines \cite{wohlin2014guidelines,jalali2012systematic}.

First, we run the search engine of DBLP for all papers containing ``test'' and ``amplification'' in their title (using stemming, which means that ``amplifying'' is matched as well).
This has resulted in 70 papers at  the date of the search (March 27, 2018)\footnote{the data is available at \url{https://github.com/STAMP-project/docs-forum/blob/master/scientific-data/}}.
We have reviewed them one by one to see whether they fit in our scope according to the definition of \autoref{sec:core-definition}. This has resulted in four articles~\cite{HamletV93,zhang2012,leung12,Joshi07}, which are the seed papers of this  literature study. The reason behind this very low proportion (4/70) is that most articles in this DBLP search are in the hardware research community, and hence do not fall in the scope of our paper.

\rev{We now briefly describe these four seed papers. More details are given in the following sections.}
\begin{itemize}
	\item \rev{\cite{HamletV93} Hamlet and Voas introduce study how different testing planning strategies can amplify testability properties of a software system.}
	\item \rev{\cite{zhang2012} Zhang and Elbaum explore a new technique to amplify a test suite for finding bugs in exception handling code. Amplification consists in triggering unexpected exceptions in sequences of API calls.}
	\item \rev{\cite{leung12} Leung et al propose to modify the test execution by using information gathered from a first test execution. The information is used to derive a formal model used to detect data races in later executions.}
	\item \rev{\cite{Joshi07} Joshi et al  try to amplify the effectiveness of testing by executing both concretely and symbolically the tests.}
\end{itemize}

From the seed papers, we have performed a backward snowballing search \rev{step} \cite{jalali2012systematic}, i.e., we have looked at all their references, going backward in the citation graph. 
Two of the authors have reviewed the papers, independently. Then, these 2 authors cross-checked the outcome of their literature review, and kept each paper for which they both \rev{agreed that it} fits the definition of test amplification (cf. \autoref{sec:core-definition}).
\rev{Then}, we have performed a forward literature search \rev{step}, using the Google scholar search engine and ``cited by'' filter, from the set of papers, in order to find the most recent contributions in this area.
\rev{A backward snowballing search step and a forward snowballing search step constitute what we call an ``iteration''.}
\rev{With each iteration, we select a set of papers for our study that we obtain through the snowballing action.}
\rev{We iterate until this set of selected paper is empty, i.e., when no paper can be kept, we stop the snowballing process in both ways: backward and forward.}

Once we had selected the  papers for our study, we distinguished 4 key approaches to amplification, which we use to  classify the literature : amplification by Adding New Tests as Variants of Existing Ones (\autoref{sec:amp_add}); Amplification by Modifying Test Execution (\autoref{sec:amp_exec}); Amplification by Synthesizing New Tests with Respect to \rev{Changes (\autoref{sec:amp_change})}; Amplification by Modifying Existing Test Code
(\autoref{sec:amp_mod}).
The missing terminological consensus mentioned previously prevented the design of a classification according to Petersen's guidelines
\cite{petersen2008systematic}.
Incrementally, we have refined the four categories by analyzing the techniques and goals in each paper.
Our methodology is as follows: we assign a work to a category if the key technique of the paper corresponds to it, per a consensus between the authors.
If no category captures the gist of the paper, we have created a new category.
If two categories are found to be closely related, we merge both categories to create a new one.
The incremental refinement of these findings led to the definition of four categories to organize this literature study.


\subsection{Novelty}

There are a number of notable surveys in software testing \cite{edvardsson1999survey,mcminn2004search,anand2013orchestrated}. However none of them is dedicated to test amplification.
For instance, we refer to Edvardsson's et al's \cite{edvardsson1999survey} and McMinn et al's \cite{mcminn2004search} articles for a survey on test generation.
Yoo and Harman have structured the work on test minimization, selection and prioritization \cite{yoo2012survey} .
In the prolific literature on symbolic execution for testing, we refer the reader to the survey of 
Păsăreanu and Visser \cite{puasuareanu2009survey}.

In general, test optimization, test selection, test prioritization, test minimization, test reduction is out of the scope of this paper.

Similarly, the work on test refactoring is related, but not in scope. In particular, the work from Van Deursen et al.~\cite{vandeursen2001refactoring,DBLP:series/springer/MoonenDZB08} and Mesaros~\cite{Meszaros2006} focuses on improving the structural and diagnosability qualities of software tests, and is a mainly manual activity. In contrast, test amplification is meant to be fully automated, as other technical amplification such as sound amplification. Its goal is also different, in that \rev{its aim is to test more effectively with regard to a a given target criterion}.

Harrold et al. \cite{harrold2008retesting} discusses the problem of ``retesting software'', where there is a section related to amplification. However, it is only a light account on the topic which is now outdated.
To our knowledge, this survey is the first survey ever dedicated to test amplification.

Yusifoğlu et al. \cite{GAROUSIYUSIFOGLU2015123} discuss the new trends in software test-code engineering, and discuss the implications for researchers and practitioners in this area. To do this, they use a systematic mapping to identify areas that require more attention.
Their work covers a larger scope than our work, since they study all software test-code engineering research, methods and empirical study, while we focus specifically on test amplification, with more depth.

\section{Amplification by Adding New Tests as Variants of Existing Ones}
\label{sec:amp_add}

The most intuitive form of test amplification is to consider an existing test suite, then generate variants of the existing test cases and add those new variants into the original test suite. We denote this kind of test amplification as $AMP_{add}$. 

\medskip
\textbf{Definition: A test amplification technique $AMP_{add}$ consists of creating new tests from existing ones \rev{to achieve a given engineering goal}. The most commonly used engineering goal is to improve coverage according to a coverage criterion.}

The works listed in this section fall into this category and have been divided according to their main engineering goal.

\subsection{Example}

In this section we present an example of $AMP_{add}$ to illustrate this category of work.
Let us consider the single Java method, presented in \autoref{lst:example}.

\begin{lstlisting}[caption={Example of a toy method},label=lst:example,float,language=java,numbers=left]
class Computer {
public void compute(int integer) {
if (integer > 2) {
return integer + 2;
} else {
return integer + 1;
}
}
}

\end{lstlisting}

This method contains an if statement. The conditional expression tests the value passed through the parameter. If the value is greater than 2, then the method returns the value plus 2, otherwise it returns the value plus 1.
Applying $AMP_{add}$ requires to have existing tests. Consider the test method in \autoref{lst:example_test_method}.
This test method ensures the behavior of the program when the parameter is lower than 2, i.e., when the else branch of the if statement is executed.

\begin{lstlisting}[caption={Example of toy test method},label=lst:example_test_method,float,language=java,numbers=left] 
@Test
public void test_compute() {
Computer computer = new Computer();
int actualValue = computer.compute(1);
assertEquals(2, actualValue);
}
\end{lstlisting}

According to this test, one can say that this program is ``poorly'' tested, since only one of the two branches is covered.
One potential goal of an $AMP_{add}$ technique is to increase this branch coverage. 

\begin{lstlisting}[caption={Example of amplified toy test method},label=lst:example_test_method_amplified,float,language=java,numbers=left] 
@Test
public void amplified_test_compute() {
Computer computer = new Computer();
int actualValue = computer.compute(3);
assertEquals(5, actualValue);
}
\end{lstlisting}

Now, an $AMP_{add}$ technique may be able to generate the amplified test method shown in \autoref{lst:example_test_method_amplified}.
The test \autoref{lst:example_test_method_amplified} is easily derivable from the existing test \autoref{lst:example_test_method} because only one literal and the assertion differ.
This new test method executes the \textit{then} branch of the if statement (see \autoref{lst:example} line 2 and 3) that was not executed before. That is to say, applying $AMP_{add}$ improves the test suite, by increasing the branch coverage of the program.

\subsection{Coverage or Mutation Score Improvement}

Baudry \textit{et al.} \cite{Baudry05a} \cite{Baudry05d} improve the mutation score of an existing test suite by generating variants of existing tests through the application of specific transformations of the test cases. They iteratively run these transformations, and propose an adaptation of genetic algorithms (GA), called a bacteriological algorithm (BA), to guide the search for test cases that kill more mutants.  
The results demonstrate the ability of search-based amplification to significantly increase the mutation score of a test suite.
They evaluated their approach on 2 case studies that are .NET classes.
The evaluation shows promising results, however the result have little external validity since only 2 classes are considered.

Tillmann and Schulte \cite{tillmann2006unit} describe a technique that can generalize existing unit tests into parameterized unit tests. The basic idea behind this technique is to refactor the unit test by replacing the concrete values that appear in the body of the test with parameters, which is achieved through symbolic execution. 
Their technique's evaluation has been conducted on 5 .NET classes.

The problem of generalizing unit tests into parameterized unit tests is also studied by Thummalapenta et al. \cite{marri2010retrofitting}. Their empirical study shows that unit test generalization can be achieved with feasible effort, and can bring the benefits of additional code coverage.
They evaluated their approach on 3 applications from 1 600 to 6 200 lines of code. The result shows an increase of the branch coverage and a slight increase of the bug detection capability of the test suite.


To improve the cost efficiency of the test generation process, Yoo and Harman \cite{yoo2012} propose a technique for augmenting the input space coverage of the existing tests with new tests. The technique is based on four transformations on numerical values in test cases, i.e., shifting ($\lambda x.x+1$ and  $\lambda x.x-1$ ) and data scaling (multiply or divide the value by 2). In addition, they employ a hill-climbing algorithm based on the number of fitness function evaluations, where a fitness is the computation of the euclidean distance between two input points in a numerical space. The empirical evaluation shows that the technique can achieve better coverage than some test generation methods which generate tests from scratch.
The approach has been evaluated  on the triangle problem.
They also evaluated their approach on two specific methods from two large and complex libraries.

To maximize code coverage, Bloem et al. \cite{6958388} propose an approach that alters existing tests to get new tests that enter new terrain, i.e., uncovered features of the program.
The approach first analyzes the coverage of existing tests, and then selects all test
cases that pass a yet uncovered branch in the target function.
Finally, the approach investigates the path conditions of the selected test cases one by one to get a new test that covers a previously
uncovered branch. To vary path conditions of existing tests, the approach uses symbolic execution and model checking techniques. A case study has shown 
that the approach can achieve 100\% branch coverage fully automatically.
They first evaluate their prototype implementation on two open source examples and then present a case
study on a real industrial program of a Java Card applet firewall.
For the real program, they applied their tool on 211 test cases, and produce 37 test cases to increase the code coverage.
The diversity of the benchmark allows to make a first generalization.

Rojas et al.~\cite{rojas2016seeding} have investigated several seeding strategies for the test generation tool Evosuite. Traditionally, Evosuite generates unit test cases from scratch. In this context, seeding consists in feeding Evosuite with initial material from which the automatic generation process can start. The authors evaluate different sources for the seed: constants in the program, dynamic values, concrete types and existing test cases. In the latter case, seeding analogizes to amplification. The experiments with 28 projects from the Apache Commons repository show a 2\% improvement of code coverage, on average, compared to a generation from scratch.
The evaluation based on Apache artifacts is stronger than most related work, because Apache artifacts are known to be complex and well tested.

Patrick and Jia \cite{Patrick201736} propose \emph{Kernel Density Adaptive Random Testing} (KD-ART) to improve the effectiveness of random testing. This technique takes advantage of run-time test execution information to generate new test inputs. It first applies \emph{Adaptive Random Testing} (ART) to generate diverse values uniformly distributed over the input space. Then, they use \emph{Kernel Density Estimation} for estimating the distribution of values found to be useful; in this case, that increases the mutation score of the test suite. KD-ART can intensify the existing values by generating inputs close to the ones observed to be more useful or diversify the current inputs by using the ART approach. The authors explore the trade-offs between diversification and intensification in a benchmark of eight C programs. They achieve an 8.5\% higher mutation score than ART for programs that have simple numeric input parameters, but their approach does not show a significant increase for programs with composite inputs. The technique is able to detect mutants 15.4 times faster than ART in average.

Instead of operating at the granularity of complete test cases, Yoshida et al. \cite{Yoshida2016} propose a novel technique for automated and
fine-grained incremental generation of unit tests through minimal augmentation  of an existing test suite. Their tool, \emph{FSX}, treats each part of existing cases, 
including the test driver, test input data, and oracles, as “test intelligence", and attempts to create tests
for uncovered test targets by copying and minimally modifying existing tests wherever possible. To achieve this, the technique uses iterative, incremental refinement of test-drivers and symbolic execution.
They evaluated \emph{FSX} using four benchmarks, from 5K to 40K lines of code. This evaluation is adequate and reveals that FSX' result can be generalized.

\subsection{Fault Detection Capability Improvement}
Starting with the source code of test cases, Harder et al. \cite{Harder03} propose an approach that dynamically generates new test cases with good fault detection ability.
A generated test case is kept only if it adds new information to the specification. 
They define ``new information'' as adding new data for mining invariants with Daikon, hence producing new or modified invariants. What is unique in the paper is the augmentation criterion: helping an invariant inference technique.
They evaluated Daikon on a benchmark of 8 C programs. These programs vary from 200 to 10K line of code. It is left to future work to evaluate the approach on a real and large software application.

Pezze et al. \cite{pezze2013} observe that method calls are used as the atoms to construct test cases for both unit and integration testing, and that most of the code in integration test cases
appears in the same or similar form in unit test cases. Based on this observation, they propose an approach which uses the information provided in unit test cases about object creation and initialization to build composite cases that focus on testing the interactions between objects. The evaluation results show that the approach can reveal new interaction faults even in well tested applications. 

Writing web tests manually is time consuming, but it gives the developers the advantage of gaining domain knowledge. In contrast, most web test generation techniques are automated and systematic, but lack the
domain knowledge required to be as effective. In light of this, Milani et al. \cite{milani2014} propose an approach which combines the advantages of the two. The approach first extracts knowledge such as event sequences and assertions from the human-written tests, and then combines the knowledge with the power of automated crawling. It has been shown that the approach can effectively improve the fault detection rate of the original test suite.
They conducted an empirical evaluation on 4 open-source and large JavaScript systems. Compared to related research, we note that it is original to consider JavaScript systems.

\subsection{Oracle Improvement}
Pacheco and Ernst implement a tool called Eclat \cite{Pacheco2005}, which aims to help the tester with the difficult task of
creating effective new test inputs with constructed oracles. Eclat first uses the execution of some available correct runs to infer an operational model of the software's operation. By making use of the established operational model, Eclat then employs a classification-guided technique to generate new test inputs. Next, Eclat reduces the number of generated inputs by selecting only those that are most likely to reveal faults. Finally, Eclat adds an oracle for each remaining test input from the operational model automatically. 
They evaluated their approach on 6 small programs. They compared Eclat's result to the result of JCrasher, a state of the art tool that has the same goal than Eclat. In their experimentation, they report that Eclat perform better than JCrasher: Eclat reveals 1.1 faults on average against 0.02 for JCrasher.

Given that some test generation techniques just generate sequences of method calls but do not contain oracles for these method calls, Fraser and Zeller \cite{fraser2011generating} propose an approach to generate parametrized unit tests containing symbolic pre- and post-conditions. Taking concrete inputs and results as inputs, the technique uses test generation and mutation to systematically generalize pre- and post-conditions. Evaluation results on five open source libraries show that the approach can successfully generalize a concrete test to a parameterized unit test, which is more general and expressive, needs fewer computation steps, and achieves a higher code coverage than the original concrete test.
They used 5 open-source and large programs to evaluate the approach. According to their observation, this technique is more expensive than simply generating unit test cases.

\subsection{Debugging Effectiveness Improvement}
Baudry \textit{et al.} \cite{Baudry:2006:ITS:1134285.1134299} propose the test-for-diagnosis criterion (TfD) to evaluate the fault localization power of a test suite, and identify an attribute called Dynamic Basic Block (DBB) to characterize this criterion. A Dynamic Basic Block (DBB) contains the set of statements that are executed by the same test cases, which implies all statements in the same DBB are indistinguishable. Using an existing test suite as a starting point, they apply a search-based algorithm to optimize the test suite with new tests so that the test-for-diagnosis criterion can be satisfied. 
They evaluated their approach on two programs: a toy program and a server that simulates business meetings over the network. These two programs are less than 2K line of code long, which can be considered as small.

R{\"o}$\beta$ler et al. \cite{robetaler2012isolating} propose BugEx, which leverages test case generation to systematically isolate failure causes. The approach takes a single failing test as input and starts generating additional passing or failing tests that are similar to the failing test. Then, the approach runs these tests and captures the differences
between these runs in terms of the observed facts that are likely related with the pass/fail outcome. Finally, these differences are statistically ranked and a ranked list of facts is produced. In addition, 
more test cases are further generated to confirm or refute the relevance of a fact. It has been shown that for six out of seven real-life bugs, the approach can accurately pinpoint important failure
explaining facts.
To evaluate BugEx, they use 7 real-life case studies from 68 to 62K lines of code. The small number of considered bugs, 7, calls for more research to improve external validity.

Yu et al. \cite{Yu2013} aim at enhancing fault localization under the scenario where no appropriate test suite is available to localize the encountered fault. They propose a mutation-oriented test case augmentation technique that is capable of generating test suites with better fault localization capabilities. The technique uses some mutation operators to iteratively mutate some existing failing tests to derive new test cases potentially useful to localize the specific encountered fault. Similarly, to increase the chance of executing the specific path during crash reproduction, Xuan et al. \cite{Xuan:2015:CRV:2786805.2803206} propose an approach based on test case mutation. The approach first selects relevant test cases based on the stack trace in the crash, followed by eliminating assertions in the selected test cases, and finally uses a set of predefined mutation operators to produce new test cases that can help to reproduce the crash. 
They evaluated MuCrash on 12 bugs for Apache Commons Collections, which is  26 KLoC of source code and 29 KLoC of test code length. The  used program is quite large and open-source which increases the confidence. but using a single subject is a threat to generalization.

\subsection{Summary}

\emph{Main achievements:}
The works discussed in this section show that adding new test cases based on existing ones can make the test generation process more targeted and cost-effective. On the one hand, the test generation process can be geared towards achieving a specific engineering goal better based on how existing tests perform with respect to the goal. For instance, new tests can be intentionally generated to cover those program elements that are not covered by existing tests. Indeed, it has been shown that tests generated in this way are effective in achieving multiple engineering goals, such as improving code coverage, fault detection ability, and debugging effectiveness. On the other hand, new test cases can be generated more cost-effectively by making use of the structure or components of the existing test cases. 

\emph{Main Challenges:}
While existing tests provide a good starting point, there are some difficulties in how to make better use of the information they contain.
First, the number of new tests synthesized from existing ones can sometimes be large and hence an effective strategy should be used to select tests that help to achieve the specific engineering goal;
the concerned works are: \cite{Baudry05a, Baudry05d, Yoshida2016}.
Second, the synthesized tests have been applied to a specific set of programs and the generalization of the related approaches could be limited. 
The concerned works are: \cite{tillmann2006unit, marri2010retrofitting, yoo2012, 6958388, Patrick201736, Harder03, Pacheco2005, Baudry:2006:ITS:1134285.1134299, robetaler2012isolating, Xuan:2015:CRV:2786805.2803206}.
\rev{Third}, some techniques have known performance issues and do not scale well: \cite{milani2014, fraser2011generating}.

\section{Amplification by Synthesizing New Tests with Respect to Changes}
\label{sec:amp_change}

Software applications are typically not tested at a single point in time; they are rather tested incrementally, along with the natural evolution of the code base: new tests are typically added together with a change or a commit~\cite{azaidmanEMSE2011,DBLP:conf/icst/ZaidmanRDD08}, to verify, for instance, that a bug has been fixed or that a new feature is correctly implemented. 
In the context of test amplification, it directly translates to the idea of synthesizing new tests \rev{as a reaction to a change}. This can be seen as a specialized form $AMP_{add}$, which considers a specific change, in addition to the existing test suite, to guide the amplification.
We call this form of test amplification $AMP_{change}$.

\medskip
\textbf{Definition: Test amplification technique $AMP_{change}$ consists of adding new tests to the current test suite, by creating new tests that cover and/or observe the effects of a change in the application code.}

We first present a series of works by Xu et al., who develop and compare two alternatives of test suite augmentation, one based on genetic algorithms and the other on concolic execution. A second subsection presents the work of a group of authors that center the attention on finding testing conditions to exercise the portions of code that exhibit changes. A third subsection exposes works that explore the adaptation and evolution of test cases to cope with code changes. The last subsection shows other promising works in this area. 

\subsection{Example}

\autoref{lst:example:ampchange:original} shows a toy class and two test cases designed to verify its code. At some point in development, the code of the method is modified as shown in \autoref{lst:example:ampchange:modified}. The change consists of the addition of a new block in line \ref{line:example:ampchange:modified}. 

\begin{lstlisting}[caption={Initial version of a class and two test cases},label=lst:example:ampchange:original,float,language=java,numbers=left]
class Computer{
public int computeValue(int input) {
if(input < 3) {
return input/2;
}
return 0;
}
}

class ComputerTest {
int threshold = 4;

@Test
public testSmallInput() {
Computer comp = new Computer();
assertTrue(comp.computeValue(2) < threshold);
}

@Test
public testDefault() {
Computer comp = new Computer();
assertEquals(comp.computeValue(10), 0);
}
}
\end{lstlisting}

\begin{lstlisting}[caption={Modified version of the initial class},label=lst:example:ampchange:modified,float,language=java,numbers=left]
class Computer{
public int computeValue(int input) {
if(input < 3) {
return input/2;
}
if (input <= 5) { (*@ \label{line:example:ampchange:modified} @*)
return 2*input;
}
return 0;
}
}
\end{lstlisting}

The existing test cases do not execute the new code. There is no test input in the $[3,5]$ interval. An $AMP_{change}$ technique would increment the test suite with a new test case, like the one shown in \autoref{lst:example:ampchange:amplified}, that covers the new code. The technique should be able to generate an input that meets the requirement to reach the new or changed code and the right oracle given the new conditions. 

\begin{lstlisting}[caption={A test case that covers the new portion of  code.},label=lst:example:ampchange:amplified,float,language=java,numbers=left]
@Test
public testInput() {
Computer comp = new Computer();
assertTrue(comp.computeValue(4) > threshold);
}
\end{lstlisting}

\subsection{Search-based vs. Concolic Approaches}

In their work, Xu et al.~\cite{xu2009directed} focus on the scenario where a program has evolved into a new version through code changes in development. They consider  techniques as (i) the identification of coverage requirements for this new version, given an existing test suite; and (ii) the creation of new test cases that exercise these requirements. Their approach first identifies the parts of the evolved program that are not covered by the existing test suite. In the same process they gather path conditions for every test case. Then, they exploit these path conditions with a concolic testing method to find new test cases for uncovered branches, analyzing one branch at a time.

\rev{Symbolic execution is a program analysis technique to reason about the execution of every path and to build a symbolic expression for each variable. Concolic testing also carries a symbolic state of the program, but overcomes some limitations of a fully symbolic execution by also considering certain concrete values. Both techniques are known to be computationally expensive for large programs.}

\rev{Xu et al. avoid a full concolic execution by only targeting paths related to uncovered branches. This improves the performance of the augmentation process.} They applied their technique to 22 versions of a small arithmetic program from the SIR \cite{SIR} repository and achieved branch coverage rates between 95\% and 100\%. They also show that a full concolic testing is not able to obtain such high coverage rates and needs a significantly higher number of constraint solver calls.

In subsequent work, Xu et al. \cite{xu2010factors} address the same problem with a genetic algorithm. Each time the algorithm runs, it targets a branch of the new program that is not yet covered. 
The fitness function measures how far a test case  falls from the target branch during its execution. The authors investigate if all test cases should be used as population, or only a subset related to the target branch or, if newly generated cases should be combined with existing ones in the population. Several variants are compared according to their \rev{cost in terms of test executions and their effectiveness, that is, whether the generated test cases achieve the goal of exercising the uncovered branches.}
The experimentation targets 3 versions of \emph{Nanoxml}, an XML parser implemented in Java with more than 7 KLoC and included in the SIR \cite{SIR} repository.
The authors conclude that considering all tests achieves the best coverage, but also requires more computational effort. They imply that the combination of new and existing test cases is an important factor to consider in practical applications. 

Xu et al. then dedicate a paper to the comparison of concolic execution and genetic algorithms for test suite amplification \cite{xu2010directed}. 
The comparison is carried out over four small (between 138 and 516 LoC) C programs from the SIR \cite{SIR} repository.
They conclude that both techniques benefit from reusing existing test cases at a cost in efficiency. The authors also state that the concolic approach can generate test cases  effectively in the absence of complex symbolic expressions. Nevertheless, the genetic algorithm is more effective in the general case, but could be more costly in test case generation. Also, the genetic approach is more flexible in terms of scenarios where it can be used, but the quality of the obtained results is heavily influenced by the definition of the fitness function, mutation test and crossover strategy. 

The same authors propose a hybrid approach \cite{xu2011hybrid}. This new approach incrementally runs both the concolic and genetic methods. Each round applies first the concolic testing and the output is passed to the genetic algorithm as initial population. Their original intention was to get a more cost-effective approach. 
The evaluation is done over three of the C programs from their previous study.
The authors conclude that this new proposal outperforms the other two in terms of branch coverage, but in the end is not more efficient. They also speculate about possible strategies for combining both individual approaches to overcome their respective weaknesses and exploit their best features. 
A revised and extended version of this work is given in \cite{xu2015directed}.

\subsection{Finding Test Conditions in the Presence of Changes}

Another group of authors have worked under the premise that achieving only coverage may not be sufficient to adequately exercise changes in code. Sometimes these changes manifest themselves only when particular conditions are met by the input. The following papers address the problem of finding concrete input conditions that not only can execute the changed code, but also propagate the effects of this change to an observable point that could be the output of the involved test cases. \rev{However, their work does not create concrete new test cases. Their goal is to provide guidance, in the form of conditions that can be leveraged to create new tests with any generation method.} 

It is important to notice that they do not achieve test generation. Their goal is to provide guidance to generate new test cases independently of the selected generation method.

Apiwattanapong et al. \cite{apiwattanapong2006matrix} target the problem of finding test conditions that could propagate the effects of a change in a program to a certain execution point. Their method takes as input two versions of the same program. First, an alignment of the statements in both versions is performed. Then, starting from the originally changed statement and its counterpart in the new version, all statements whose execution is affected by the change are gathered up to a certain distance. The distance is computed over the control and data dependency graph.  A partial symbolic execution is performed over the affected instructions to retrieve the states of both program versions, which are in turn used to compute testing requirements that can propagate the effects of the original change to the given distance. As said before, the method does not deal with test case creation, it only finds  new testing conditions that could be used in a separate generation process and is not able to handle changes to several statements unless the changed statements are unrelated. 
The approach is evaluated on Java translations of two small C programs (102 Loc and 268 LoC) originally included in the Siemens program dataset \cite{hutchins1994experiments}. The authors conclude that, although limited to one change at a time, the technique can be leveraged to generate new test cases during regular development.

Santelices et al. \cite{santelices2008test} continue and extend the previous work by addressing changes to multiple statements and considering the effects they could have on each other. In order to achieve this they do not compute state requirements for changes affected by others. This time, the evaluation is done in one of the study subjects form their previous study and two versions of \emph{Nanoxml} from SIR. 

In another paper \cite{santelices2011applying} the same authors address the problems in terms of efficiency of applying symbolic execution. They state that limiting the analysis of affected statements up to a certain distance from changes reduces the computational cost, but scalability issues still exist. They also explain that their previous approach often produces test conditions which are unfeasible or difficult to satisfy within a reasonable resource budget. To overcome this, they perform a dynamic inspection of the program during test case execution over statically computed slices around changes. The technique is evaluated over five small Java programs, comprising \emph{Nanoxml} with 3 KLoC and translations of C programs from SIR having between 283 LoC and 478 LoC. This approach also considers multiple program changes. Removing the need of symbolic execution leads to a less expensive method. The authors claim that propagation-based testing strategies are superior to coverage-based in the presence of evolving software.

\subsection{Other Approaches}

Other authors have also explored test suite augmentation for evolving programs with propagation-based approaches. Qui et al. \cite{qi2010test} propose a method to add new test cases to an existing test suite ensuring that the effects of changes in the new program version are observed in the test output. The technique consists of a two step symbolic execution. First, they explore the paths towards a change in the program guided by a notion of distance over the control dependency graph. This exploration produces an input able to reach the change. In a second moment they analyze the conditions under which this input may affect the output and make changes to the input accordingly. The technique is evaluated using 41 versions of the \emph{tcas} program from the SIR repository (179 LoC) with only one change between versions. The approach was able to generate tests reaching the changes and affected the program output for 39 of the cases. Another evaluation was also included for two consecutive versions of the \emph{libPNG} library (28 KLoC) with a total of 10 independent changes between them. The proposed technique was able to generate tests that reached the changes in all cases and the output was affected in nine of the changes. The authors conclude that the technique is effective in the generation of test inputs to reach a change in the code and expose the change in the program output.

Wang et al. \cite{xwang2014directed} exploit existing test cases to generate new ones that execute the change in the program. These new test cases should produce a new program state, in terms of variable values, that can be propagated to the test output. An existing test case is analyzed to check if it can reach the change in an evolved program. The test is also checked to see if it produces a different program state at some point and if the test output is affected by the change. If some of these premises do not hold then the path condition of the test is used to generate a new path condition to achieve the three goals. Further path exploration is guided and narrowed using a notion of the probability for the path condition to reach the change. This probability is computed using the distance between statements over the control dependency graph. Practical results of test cases generation in three small Java programs (from 231 LoC to 375 LoC) are exhibited. The method is compared to \emph{eXpress} and \emph{JPF-SE} two state of the art tools and is shown to reduce the number of symbolic executions by 45.6\% and 60.1\% respectively. As drawback, the technique is not able to deal with changes on more than one statement. 

Mirzaaghaei \textit{et al.} \cite{Mirzaaghaei2012,mirzaaghaei2014automatic} introduce an approach that leverages information from existing test cases and automatically adapts test suites to code changes. Their technique can repair, or evolve test cases in front of signature changes (\textit{i.e.} changing the declaration of method parameters or return values), the addition of new classes to the hierarchy, addition of new interface implementations, new method overloads and new method overrides. Their effective implementation \emph{TestCareAssitance} (TCA) first diffs the original program with its modified version to detect changes and searches in the test code similar patterns that could be used to complete the missing information or change the existing code. They evaluate TCA for signature changes in 9 Java projects of the Apache foundation and repair in average 45\% of modifications that lead to compilation errors. The authors further use five additional open source projects to evaluate their approach when adding new classes to the hierarchy. TCA is able to generate test cases for 60\% of the newly added classes.
This proposal could also fall in the category of test repairing techniques. Section \ref{sec:amp_mod} will explore alternatives in a similar direction that produce test changes instead of creating completely new test cases.

In a different direction, Böhme et al. \cite{bohme2013regression} explain that changes in a program should not be treated in isolation. Their proposal focuses on potential interaction errors between software changes. They propose to build a graph containing the relationship between changed statements in two different versions of a program and potential interaction locations according to data and control dependency. This graph is used to guide a symbolic execution method and find path conditions for exercising changes and their potential interactions and use a Satisfiability Modulo Solver to generate a concrete test input. They provide practical results on six versions the \emph{GNU Coreutils} toolset that introduce 11 known errors. They were able to find 5 unknown errors in addition to previously reported issues.


Marinescu and Cadar \cite{marinescu2013katch} present a system, called \emph{Katch}, that aims at covering the code included in a patch. Instead of dealing with one change to one statement, as most of the previous works, this approach first determines the differences of a program and its previous version after a commit, in the form of a code patch. Lines included in the patch are filtered by removing those that contain non-executable code (i.e., comments, declarations). If several lines belong to the same basic program block, only one of them is kept as they will all be executed together. From the filtered set of lines, those not covered by the existing test suite are considered as targets. The approach then selects the closest input to each target from existing tests using the static minimum distance over the control flow graph. Edges on this graph that render the target unreachable are removed by inspecting the data flow and gathering preconditions to the execution of basic blocks. To generate new test inputs, they combine symbolic execution with heuristics that select branches by their distance to the target, regenerate a path by going back to the point where the condition became unfeasible or changing the definition of variables involved in the condition. The proposal is evaluated using the \emph{GNU findutils}, \emph{diffutils} and \emph{binutils} which are distributed with  most Unix-based distributions. They examine patches from a period of 3 years. In average, they automatically increase coverage from 35\% to 52\% with respect to the manually written test suite.

A posterior work of the same group~\cite{palikareva2016shadow} also targets patches of code, focusing on finding test inputs that execute different behavior between two program versions. They consider two versions of the same program, or the old version with the patch of changed code, and a test suite. The code should be annotated in places where changes occur in order to unify both versions of the program for the next steps. Then they select from the test suite those test cases that cover the changed code. If there is no such test case, it can be generated using \emph{Katch}.  
The unified program is used in a two stage dynamic symbolic execution guided by the selected test cases: look for branch points where two semantically different conditions are evaluated in both program versions; bounded symbolic execution for each point previously detected. At those points all possible alternatives in which program versions execute the same or different branch blocks are considered and used to make the constraint solver generate new test inputs for divergent scenarios. The program versions are then normally executed with the generated inputs and the result is validated to check the presence of a bug or an intended difference. In their experiments this validation is mostly automatic but in general should be performed by developers. The evaluation of the proposed method is based on the \emph{CoREBench}~\cite{bohme2014corebench} data set that contains documented bugs and patches of the \emph{GNU Coreutils} program suite. The authors discuss successful and unsuccessful results but in general the tool is able to produce test inputs that reveal changes in program behaviour.

\subsection{Summary}

\emph{Main achievements:}
$AMP_{change}$ techniques often rely on symbolic and concolic execution. Both have been successfully combined with other techniques in order to generate test cases that reach changed or evolved parts of a program \cite{xu2011hybrid,xu2015directed,marinescu2013katch}. Those hybrid approaches produce new test inputs that increase the coverage of the new program version. Data and control dependency has been used in several approaches to guide symbolic execution and reduce its computational cost \cite{bohme2013regression,marinescu2013katch,xwang2014directed}. The notion of distance from statements to observed changes has been also used for this matter \cite{marinescu2013katch,apiwattanapong2006matrix}.

\emph{Main challenges:}
Despite the progress made in the area, a number of challenges remain open. The main challenge relates to the size of the changes considered for test amplification: many of the works in this area consider a single change in a single statement \cite{apiwattanapong2006matrix,qi2010test,xwang2014directed}. While this is relevant and important to establish the foundations for $AMP_{change}$, this cannot fit current development practices where a change, usually a commit, modifies the code at multiple places at once. A few papers have started investigating multi-statement changes for test suite amplification \cite{santelices2008test,marinescu2013katch,palikareva2016shadow}. Now,  $AMP_{change}$  techniques should fit into the revision process and be able to consider a commit as the unit of change. 

Another challenge relates to scalability. The use of symbolic and concolic execution has proven to be effective in test input generation targeting program changes. Yet,  these two techniques are computationally expensive \cite{xu2009directed,xu2011hybrid,xu2015directed,apiwattanapong2006matrix,santelices2008test,palikareva2016shadow}. Future works shall consider more efficient ways for exploring input requirements that  exercise program changes or new uncovered parts. Santelices and Harrold~\cite{santelices2011applying} propose to get rid of symbolic execution by observing the program behavior during test execution. However, they do not generate test cases.

Practical experimentation and evaluation remains confined to a very small number of programs, in most cases less than five study subjects, and even small programs in terms of effective lines of code. A large scale study on the subject is still missing.

\section{Amplification by Modifying Test Execution}
\label{sec:amp_exec}

In order to explore new program states and behavior, it is possible to interfere with the execution at runtime so as to modify the execution of the program under test. 

\medskip
\textbf{Definition: Test amplification technique $AMP_{exec}$ consists of modifying the test execution process or the test harness in order to maximize the knowledge gained from the testing process.}

\rev{One of the drawbacks of automated tests is the hidden dependencies that may exist between different unit test cases. In fact, the order in which the test cases are executed may affect the state of the program under test. A good and strong test suite should have no implicit dependencies between test cases.}

The majority of test frameworks are deterministic, i.e., between two runs the order of execution of test is the same~\cite{DBLP:conf/icsm/PalombaZ17,PalombaEMSE2019}.

\rev{An $AMP_{exec}$ technique would randomize the order in which the tests are executed to reveal hidden dependencies between unit tests and potential bugs derived from this situation.}

\subsection{Amplification by Modifying Test Execution}

Zhang and Elbaum \cite{zhang2012,ZhangE14} describe a technique to validate exception handling in programs making use of APIs to access external resources such as databases, GPS or bluetooth. The method mocks the accessed resources and amplifies the test suite by triggering unexpected exceptions in sequences of API calls. Issues are detected during testing by observing abnormal terminations of the program or abnormal execution times. 
They evaluated their approach on 5 Android artifacts. Their sizes vary from 6k to 18k line of codes, with 39 to 117 unit tests in the test suite. The size of the benchmark seems quite reasonable. The approach is shown to be cost-effective and able to detect real-life problems in 5 Android applications.

Cornu et al. \cite{cornu2015exception} work in the same line of exception handling evaluation. They propose a method to complement a test suite in order to check the behaviour of a program in the presence of unanticipated scenarios. The original code of the program is modified with the insertion of \texttt{throw} instructions inside \texttt{try} blocks. \rev{The test suite is considered as an executable specification of the program and therefore used as an oracle in order to compare the program execution before and after the modification.} Under certain conditions, issues can be automatically repaired by catch-stretching.
\rev{The authors used nine Java open-source projects to create a benchmark and evaluate their approach.} This benchmark is big enough to conclude the generalization of the results. The selected artifacts are well-known, modern and large: Apache artifacts, joda-time and so on.
Their empirical evaluation shows that the short-circuit testing approach of exception contracts increases the knowledge of software.

Leung et al \cite{leung12} are interested in finding \rev{data races} and non-determinism in GPU code written in the CUDA programming language. 
In their context, test amplification consists of generalizing the information learned from a single dynamic run. 
The main contribution is to formalize the relationship between the trace of the dynamic run and statically collected information flow. 
The authors leverage this formal model to define the conditions under which they can generalize the absence of race conditions for a set of input values, starting from a run of the program with a single input.
They evaluated their approach \rev{using} 28 benchmarks in the
NVIDIA CUDA SDK Version 3.0. They removed trivial ones and some of them that they cannot handle. The set of benchmarks is big enough and contains a diversity of applications to be convinced that the approach can be generalized.

Fang et al. \cite{fang2015perfblower} develop a performance testing system named \emph{Perfblower}, which is able to detect and diagnose memory issues by observing the execution of a set of test cases. The system includes a domain-specific language designed to describe memory usage symptoms. Based on the provided descriptions, the tool evaluates the presence of memory problems. The approach is evaluated on 13 Java real-life projects. The tool is able to find real memory issues and reduce the number of false positives reported \rev{by similar tools}.
They used the small workload of the DaCapo~\cite{DaCapo} benchmark. \rev{They argue that developers will not use large workloads and it is much more difficult to reveal performance bugs under small workloads.} These two claims are legit, \rev{however the authors do not provide} any evidence of the scalability of the approach.

Zhang et al. \cite{Zhang2016Isomorphic} devise a methodology to improve the capacity of the test suite to detect regression faults. Their approach is able to exercise uncovered branches without generating new test cases. They first look for identical code fragments between a program and its previous version. Then, new variants of both versions are generated by negating branch conditions that force the test suite to execute originally uncovered parts. The behaviour of version variants are compared through test outputs. An observed difference in the output could reveal an undetected fault. An implementation of the approach is compared with \emph{EvoSuite} \cite{fraser2011evosuite} in 10 real-life Java projects. In the experiments known faults are seeded by mutating the original program code. The results show that \emph{EvoSuite} obtains better branch coverage, while the proposed method is able to detect more faults. The implementation is available in the form of a tool named \emph{Ison}.

\subsection{Summary}

\emph{Main achievements:} \rev{$AMP_{exec}$ proposals} provide cost-effective approaches to observe and modify a program execution to detect possible faults. This is done by instrumenting the original program code to place observations at certain points or mocking resources to monitor API calls and explore unexpected scenarios. It adds no prohibitive overheads to regular test execution and provides means to gather useful runtime information. 
Techniques in this section were used to analyze real-life projects of different sizes and they are shown to match other tools that pursue the same goal and obtain better results in some cases. 

\emph{Main challenges:} 
As shown by the relatively small number of papers discussed in this section, \rev{techniques for test execution modification have not been widely explored}. The main challenge is to get this concept known so as to enlarge the research community working on this topic.
The concerned works are: \cite{zhang2012,ZhangE14,cornu2015exception,leung12,fang2015perfblower,Zhang2016Isomorphic}.

\section{Amplification by Modifying Existing Test Code}
\label{sec:amp_mod}

In testing, it is up to the developer to design integration (large) or unit (small) tests. The main testing infrastructure such as JUnit in Java does not impose anything on the tests, such as the number of statements in a test, the cohesion of test assertions or the meaningfulness of test methods grouped in a test class. \rev{In literature}, there is work on modifying existing tests with respect to a certain engineering goal.

\medskip
\textbf{Definition: Test amplification technique $AMP_{mod}$ refers to modifying the body of existing test methods. The goal here is to make the scope of each test cases more precise or to improve the ability of test cases at assessing correctness (with better oracles). Differently from $AMP_{add}$, it is not about adding new test methods or new tests classes.}

\subsection{Example}

\begin{lstlisting}[caption={Example of a toy class},label=lst:example:amplification:original,float,language=java,numbers=left]
public class Stack {
private Comparable[] elems;
public Stack() { ... }
public void push(Comparable i) { ... }
public void pop() { ... }
public boolean isFull() { ... }
public boolean isEmpty() { ... }
}
\end{lstlisting}

\begin{lstlisting}[caption={Initial test suite for the toy class},label=lst:example:test:initial,float,language=java,numbers=left]
public class StackTest {
@Test
public void test1() {  
Stack s1 = new Stack();
s1.push('a');
s1.pop();
}
}
\end{lstlisting}

\begin{lstlisting}[caption={Augmented test suite for the toy class},label=lst:example:test:aug,float,language=java,numbers=left]
public  class  StackTest {
@Test
public  void  testAug1 () {
Stack s1 = new  Stack();
assertTrue(s1.isEmpty());
assertFalse(s1.isFull());
s1.push('a');
assertFalse(s1.isEmpty());
assertFalse(s1.isFull());
s1.pop();
}
}
\end{lstlisting}

We now use an example to give an illustration of work in this category. Let us consider a simple Java class named \emph{Stack} in \autoref{lst:example:amplification:original}. The example is a simplified Java implementation of a stack that stores unique elements. In the implementation, the array \emph{elems} contains the elements of the stack, and the \emph{push} and \emph{pop} functions represent the two standard push and pop stack operations. The functions \emph{isFull} and \emph{isEmpty} check whether the stack is full and empty respectively.

Given the Java class, existing automatic test-generation tools can generate a test suite for it. For instance, \autoref{lst:example:test:initial} exemplifies a possible test generated by automatic test-generation tools. Note however there are no assertions generated in the test suite. To detect problems during test execution, it typically relies on observing whether uncaught exceptions are thrown or whether the execution violates some predefined contracts. 

A test amplification technique $AMP_{mod}$ may be able to generate the amplified test suite as shown in \autoref{lst:example:test:aug}.
Compared with the original test suite, the augmented test suite has comprehensive assertions. These assertions reflect the behavior of the current program version under test and can be used to detect regression faults introduced in future program versions.

\subsection{Input Space Exploration}
Dallmeier et al. \cite{Dallmeier2010} automatically amplify test suites by adding and removing method calls in JUnit test cases. Their objective is to produce test cases that cover a wider set of executions than the original test suite in order to improve the quality of models reverse engineered from the code. They evaluate \emph{TAUTOKO} on 7 Java classes and \rev{show that it is able to produce} richer typestates (a typestate is a finite state automaton which encodes legal usages of a class under test).

Hamlet and Voas \cite{HamletV93} introduce the notion of ``reliability amplification'' to establish a better statistical confidence that a given software is correct. Program reliability is measured as the mean time to failure of the system under test. The core contribution relates reliability to testability assessment, that is, a measure of the probability that a fault in the program will propagate to an observable state. The authors discuss how different systematic test planning strategies, e.g., partition-based test selection \cite{ostrand1988category}, can complement profile-based test cases, in order to obtain a better measurement of testability and therefore better bounds to estimate the reliability of the program being tested.

\subsection{Oracle Improvement}

Xie \cite{Xie2006} amplifies object-oriented unit tests with a the technique that consists of adding assertions on the state of the receiver object, the returned value by the tested method (if it is a non-void return value method) and the state of parameters (if they are not primitive values). Those values depend on the behavior of the given method, which in turn depends on the state of the receiver and of arguments at the beginning of the invocation. The approach, named \emph{Orstra}, consists of instrumenting the code and running the test suite to collect state of objects. Then, assertions are generated, which call observer methods (methods with a non-void return type, \textit{e.g.}, $toString()$). To evaluate \emph{Orstra}, the author uses 11 Java classes from a variety of sources. Theses classes are different in the number of methods and lines of code, and the author also uses two different third-party test generation tools to generate the initial test suite \rev{to be augmented}. The results show that \emph{Orstra} can effectively improve the fault-detection capability of the original automatically generated test suite.

Carzaniga \textit{et al.} \cite{Carzaniga:2014:COI:2568225.2568287} reason about generic oracles and propose a generic procedure to assert the behavior of a system under test. To do so, they exploit the redundancy of software. Redundancy of software happens when the system can perform the same action through different executions, either with different code or with the same code but with different input parameters or in different contexts. They devise the notion of ``cross-checking oracles'', which compare the outcome of the execution of an original method to the outcome of an equivalent method. Such oracle uses a generic equivalence check on the returned values and the state of the target object. If there is an inconsistency, the oracle reports it, otherwise, the checking continue. These oracles are added to an existing test suite with aspect-oriented programming. For the evaluation, they use 18 classes from three non-trivial open-source Java libraries, including Guava, Joda-Time, and GraphStream. The subject classes are selected based on whether a set of equivalences have already been established or could be identified. For each subject class, two kinds of test suites have been used, including hand-written test suites and automatically generated test suites by Randoop. The experimental results show that the approach can slightly increase (+6\% overall) the mutation score of a manual test suite.

Joshi et al. \cite{Joshi07} try to amplify the effectiveness of testing by executing both concretely and symbolically the tests. Along this double execution, for every conditional statement executed by the concrete execution, the symbolic execution generates symbolic constraints over the input variables. At the execution of an assertion, the symbolic execution engine invokes a theorem prover to check that the assertion is verified, according to the constraints encountered. If the assertion is not guaranteed, a violation of the behavior is reported. To evaluate their approach, the authors use 5 small and medium sized programs from SIR, including gzip, bc, hoc, space, and printtokens. The results show that they are able to detect buffer overflows but it needs optimization because of the huge overhead that the instrumentation add.

Mouelhi \textit{et al.} \cite{Mouelhi:2009} enhance tests oracles for access  control  logic, also called Policy Decision Point (PDP). This is done in 3 steps: select test cases that execute PDPs, map each of the test cases to specific PDPs and  oracle enhancement. They add to the existing oracle checks that the access is granted or denied with respect to the rule and checks that the PDP is correctly called. To do so, they force the Policy Enforcement Point, \textit{i.e.}, the point where the policy decision is setting in the system functionality, to raise an exception when the access is denied and they compare the produced logs with expected log. To evaluate, they conduct case studies on three Java applications developed by students during group projects. For these three subjects, the number of classes ranges form 62 to 122,  the number of methods ranges from 335 to 797,  and the number of lines of code ranges from 3204 to 10703. The experimental results show that compared to manual testing, automated oracle generation saves a lot of time (from 32 hours to 5 minutes).

Daniel \emph{et al.} \cite{reassert2009} devise \emph{ReAssert} to  automatically repair test cases, i.e., to modify test cases that fail due to a change. \emph{ReAssert} follows five steps: record the values of failing assertions, re-executes the test and catch the failure exception, \textit{i.e.}, the exception thrown by the failing assertion. From the exception, it extracts the stack trace to find the code to repair. Then, it selects the repair strategy depending on the structure of the code and on the recorded value. Finally, ReAssert re-compiles the code changes and repeats all steps until no more assertions fail. The tool was evaluated on six real and well known open source Java projects, namely \emph{PMD}, \emph{JFreeChart}, \emph{Lucene}, \emph{Checkstyle}, \emph{JDepend} and \emph{XStream}. The authors created a collection of manually written and generated tests cases by targeting previous versions of these programs. \emph{ReAssert} was able to produce fixes from 25\% to 100\% of failing tests for all study subjects. An usability study was also carried out with two teams of 18 researchers working on three research prototypes. The participants were asked to accomplish a number of tasks to write failing tests for new requirements and code changes and were also asked to manually fix the failures. \emph{ReAssert} could repair 98\% of failures created by the participants’ code changes. In 90 \% of cases the repairs suggested by the tool  matched the patches created by the participants. The authors explain that the success rate of the tool depends more on the structure of the code of the test than the test failure itself.

\subsection{Purification}

Xuan et al. \cite{xuan:hal-01309004} propose a technique to split existing tests into smaller parts in order to ``purify'' test cases. Here, purification can be seen as a form of test refactoring. A pure test executes one, and only one, branch of an if/then/else statement. On the contrary, an impure test executes both branches \emph{then} and \emph{else} of the same if/then/else statement in code. \rev{The authors evaluate} their technique on 5 widely used open-source projects from code organizations such as Apache. The experimental results show that the technique increases the purity of test cases by up to 66\% for if statements and 11\% for try statement. In addition, the result also shows that the technique improves the effectiveness of program repair of Nopol~\cite{xuanTSE2016Nopol}.

Xuan \textit{et al.} \cite{xuan2014test} aim at improving the fault localization capabilities by \emph{purifying} test cases. By purifying, they mean to modify existing failing test cases into single assertion test cases and remove all statements that are not related to the assertion. They evaluated the test purification on 6 open-source java project, over 1800 bugs generated by typical mutation tool PIT and compare their results with 6 mature fault localization techniques. They show that they improve the fault localization effectiveness on 18 to 43\% of all the faults, as measured per improved wasted effort. 





\subsection{Summary}

\emph{Main achievements:}
What is remarkable in $AMP_{mod}$ is the diversity of engineering goals considered. 
Input space exploration provides better state coverage \cite{Dallmeier2010} and reliability assessment \cite{HamletV93}, 
oracle improvement \rev{allows to increase the efficiency and effectiveness of tests} \cite{Xie2006, Carzaniga:2014:COI:2568225.2568287, Joshi07, Mouelhi:2009, reassert2009}, test purification of test cases facilitate program repair \cite{xuan:hal-01309004} and fault localization \cite{xuan2014test}.

\emph{Main challenges:}
Although impressive results have been obtained, no experiments have been carried out to study the acceptability and maintainability of amplified tests \cite{Dallmeier2010, Xie2006, HamletV93,  Carzaniga:2014:COI:2568225.2568287, Joshi07, Mouelhi:2009, reassert2009, xuan:hal-01309004, xuan2014test}. In this context, acceptability means that human developers are ready to commit the amplified tests to the version control system (e.g., the Git repository). The maintainability challenge is whether the machine-generated tests can be later understood and modified by developers. \rev{To our understanding}, these are the main challenges of test code modification.

\section{Analysis}
\label{sec:analysis}

\rev{ \autoref{tab:snowballing_steps} shows a summary of the results of our snowballing review. Every row corresponds to one iteration in the process. Column \textbf{\# F-ref} shows the number of papers added by following forward references. Column \textbf{\# B-ref} shows the number of papers added by following backward references. The last column shows the total number of papers added for each iteration.} 

\begin{table}[H]
	\centering
	\begin{tabular}{c|rrr}
		Step  & \# F-ref & \# B-ref & Total \\
		\hline
		Seed  &        0 &        0 &     4 \\
		It 1  &        4 &        1 &     5 \\
		It 2  &        0 &        3 &     3 \\
		It 3  &       16 &        2 &    18 \\
		It 4  &        6 &        2 &     8 \\
		It 5  &        6 &        3 &     9 \\
		It 6  &        2 &        0 &     2 \\
		\hline
		Total &       34 &       11 &    49 \\
	\end{tabular}
	\caption{Details on the snowballing review. The first column shows the iteration, the second column contains the number of forward reference retained at a given iteration, the third column shows the number of backward reference retained at a given iteration, and the fourth column contains the total number of reference retained at a given iteration. The first row corresponds to the starting set of reference, i.e., the seed references.}
	\label{tab:snowballing_steps}
\end{table}

We now provide a recapitulation of all the dimensions considered in our study. This section provides an overall view on these papers, so that the reader can have a quick summary \rev{of the main lines of research that we analyzed}. 

\subsection{Aggregated View}

\begin{table*}
	\caption{List of papers included in this snowballing survey. The columns correspond to the article categorization, the engineering goals, techniques employed, the programming language of the systems under test and the publication details. \rev{The last column shows the iteration the paper was included in our study.}}
	
	\label{tab:summary}
	\begin{adjustbox}{max width=\textwidth,center=\textwidth}
		\rowcolors{2}{gray!25}{white}
		\begin{tabular}{l|l|llll|llllll|llllll|l|llll|l}
			\toprule
			\vertical{Reference}
			& Term used 
			& \vertical{Add new tests} 
			& \vertical{With respect to change/diff}
			& \vertical{Runtime modification}
			& \vertical{Modifies existing tests}
			& \vertical{Improve coverage}
			& \vertical{Reproduce crashes}
			& \vertical{Detect new faults}
			& \vertical{Localize faults}
			& \vertical{Improve repair}
			& \vertical{Improve observability}
			& \vertical{Test code analysis}
			& \vertical{Application code analysis}
			& \vertical{Execution modification}
			& \vertical{Concolic execution}
			& \vertical{Symbolic execution}
			& \vertical{Search based heuristics}
			& \vertical{Target language}
			& Venue
			& \vertical{Publication year} 
			& \vertical{Last name of first author} 
			& \vertical{Iteration} \\
			\midrule
			\cite{Harder03}                                 & augmentation                      & \X &    &    &    &    &    & \X &    &    &    & \X &    &    &    &    &    & C           & ICSE                                                     & 2003 & Harder         & 2 \\
			\cite{Baudry:2002:ATC:786769.787015}            & optimization                      & \X &    &    &    & \X &    &    &    &    &    & \X &    &    &    &    & \X & .NET        & ASE                                                      & 2002 & Baudry         & 2 \\
			\cite{Baudry05a}                                & optimization                      & \X &    &    &    & \X &    &    &    &    &    & \X &    &    &    &    & \X & Eiffel, C\# & STVR                                                     & 2005 & Baudry         & 3 \\
			\cite{Baudry05d}                                & optimization                      & \X &    &    &    & \X &    &    &    &    &    & \X &    &    &    &    & \X & C\#         & IEEE Software                                            & 2005 & Baudry         & 3 \\
			\cite{Pacheco2005}                              & generation                        & \X &    &    &    &    &    & \X &    &    &    & \X &    &    &    &    &    & Java        & ECOOP                                                    & 2005 & Pacheco        & 1 \\
			\cite{Baudry:2006:ITS:1134285.1134299}          & optimization                      & \X &    &    &    &    &    &    & \X &    &    & \X &    &    &    &    &    & Java        & ICSE                                                     & 2006 & Baudry         & 4 \\
			\cite{tillmann2006unit}                         & generation                        & \X &    &    & \X & \X &    &    &    &    &    & \X &    &    &    & \X &    & Spec\#      & IEEE Software                                            & 2006 & Tillmann       & 5 \\
			\cite{marri2010retrofitting}                    & generalization                    & \X &    &    & \X & \X &    &    &    &    &    & \X &    &    &    & \X &    & C\#         & FASE                                                     & 2011 & Thummalapenta  & 5 \\
			\cite{fraser2011generating}                     & generation                        & \X &    &    &    &    &    & \X &    &    &    & \X & \X &    &    &    &    & Java        & ISSTA                                                    & 2011 & Fraser         & 6 \\
			\cite{robetaler2012isolating}                   & generation                        & \X &    &    &    &    &    &    & \X &    &    & \X &    &    &    &    &    & Java        & ISSTA                                                    & 2012 & Ropler         & 5 \\
			\cite{yoo2012}                                  & regeneration                      & \X &    &    &    & \X &    &    &    &    &    & \X &    &    &    &    & \X & Java        & STVR                                                     & 2012 & Yoo            & 4 \\
			\cite{pezze2013}                                & generation                        & \X &    &    &    &    &    & \X &    &    &    & \X &    &    &    &    &    & Java        & ICST                                                     & 2013 & Pezze          & 5 \\
			\cite{Yu2013}                                   & augmentation                      & \X &    &    &    &    &    &    & \X &    &    & \X &    &    &    &    &    & Java        & IST                                                      & 2013 & Yu             & 5 \\
			\cite{6958388}                                  & augmentation                      & \X &    &    &    & \X &    &    &    &    &    & \X &    &    &    & \X &    & C           & QSIC                                                     & 2014 & Bloem          & 3 \\
			\cite{milani2014}                               & generation                        & \X &    &    &    &    &    & \X &    &    &    & \X &    &    &    &    &    & JavaScript  & ASE                                                      & 2014 & Fard           & 3 \\
			\cite{Xuan:2015:CRV:2786805.2803206}            & mutation                          & \X &    &    &    &    & \X &    &    &    &    & \X &    &    &    &    &    & Java        & ESEC/FSE                                                 & 2015 & Xuan           & 3 \\
			\cite{rojas2016seeding}                         & generation                        & \X &    &    &    & \X &    &    &    &    &    & \X &    &    &    &    & \X & Java        & STVR                                                     & 2016 & Rojas          & 5 \\
			\cite{Yoshida2016}                              & augmentation                      & \X &    &    & \X & \X &    &    &    &    &    & \X &    &    &    & \X &    & C, C++      & ISSTA                                                    & 2016 & Yoshida        & 3 \\
			\cite{Patrick201736}                            & generation                        & \X &    &    &    &    &    & \X &    &    &    &    &    &    &    &    & \X & C           & IST                                                      & 2017 & Patrick        & 4 \\
			\cite{apiwattanapong2006matrix}                 & augmentation                      &    & \X &    &    & \X &    &    &    &    & \X &    & \X &    &    & \X &    & Java        & TAIC PART                                                & 2006 & Apiwattanapong & 3 \\
			\cite{santelices2008test}                       & augmentation                      &    & \X &    &    & \X &    &    &    &    & \X &    & \X &    &    & \X &    & Java        & ASE                                                      & 2008 & Santelices     & 3 \\
			\cite{reassert2009}                             & \stack{repairing}{refactoring}    &    & \X &    & \X &    &    &    &    &    & \X & \X &    &    &    &    &    & Java        & ASE                                                      & 2009 & Daniel         & 4 \\
			\cite{xu2009directed}                           & augmentation                      &    & \X &    &    & \X &    &    &    &    &    &    &    &    & \X &    &    & Java        & APSEC                                                    & 2009 & Xu             & 3 \\
			\cite{qi2010test}                               &                                   &    & \X &    &    & \X &    &    &    &    &    &    & \X &    &    & \X &    & C           & ASE                                                      & 2010 & Qi             & 4 \\
			\cite{xu2010factors}                            & augmentation                      &    & \X &    &    & \X &    &    &    &    &    &    & \X &    &    &    & \X & Java        & GECCO                                                    & 2010 & Xu             & 3 \\
			\cite{xu2010directed}                           & augmentation                      &    & \X &    &    & \X &    &    &    &    &    &    & \X &    & \X &    & \X & C           & FSE                                                      & 2010 & Xu             & 2 \\
			\cite{santelices2011applying}                   & augmentation                      &    & \X &    &    & \X &    &    &    &    & \X &    & \X &    &    &    &    & Java        & ICST                                                     & 2011 & Santelices     & 3 \\
			\cite{xu2011hybrid}                             & augmentation                      &    & \X &    &    & \X &    &    &    &    &    &    & \X &    & \X &    & \X & C           & ISSRE                                                    & 2011 & Xu             & 3 \\
			\cite{Mirzaaghaei2012}                          & \stack{repairing}{adaptation}     &    & \X &    & \X & \X &    &    &    & \X &    & \X & \X &    &    &    &    & Java        & ICST                                                     & 2012 & Mirzaaghaei    & 3 \\
			\cite{mirzaaghaei2014automatic}                 & \stack{repairing}{adaptation}     &    & \X &    & \X & \X &    &    &    & \X &    & \X & \X &    &    &    &    & Java        & SVTR                                                     & 2014 & Mirzaaghaei    & 3 \\
			\cite{bohme2013regression}                      &                                   &    & \X &    &    & \X &    & \X &    &    &    &    & \X &    &    & \X & \X & C           & ESEC/FSE                                                 & 2013 & B\"ohme        & 3 \\
			\cite{marinescu2013katch}                       &                                   &    & \X &    &    & \X &    &    &    &    &    &    & \X &    &    & \X & \X & C           & ESEC/FSE                                                 & 2013 & Marinescu      & 5 \\
			\cite{xwang2014directed}                        & augmentation                      &    & \X &    &    & \X &    &    &    &    & \X &    & \X &    &    & \X &    & Java        & CSTVA                                                    & 2014 & Wang           & 3 \\
			\cite{xu2015directed}                           & augmentation                      &    & \X &    &    & \X &    &    &    &    &    &    & \X &    & \X &    & \X & C           & STVR                                                     & 2015 & Xu             & 3 \\
			\cite{palikareva2016shadow}                     &                                   &    & \X &    &    & \X &    &    &    &    & \X &    & \X &    &    & \X &    & C           & ICSE                                                     & 2016 & Palikareva     & 4 \\
			\cite{zhang2012}                                & amplification                     &    &    & \X &    &    &    & \X &    &    &    &    & \X & \X &    &    &    & Java        & ICSE                                                     & 2012 & Zhang          & S \\
			\cite{ZhangE14}                                 & amplification                     &    &    & \X &    &    &    & \X &    &    &    &    & \X & \X &    &    &    & Java        & TOSEM                                                    & 2014 & Zhang          & \\
			\cite{leung12}                                  & amplification                     &    &    & \X &    &    &    & \X &    &    &    &    & \X &    &    &    &    & CUDA        & PLDI                                                     & 2012 & Leung          & S \\
			\cite{cornu2015exception}                       & amplification                     &    &    & \X &    &    &    & \X &    & \X &    &    & \X & \X &    &    &    & Java        & IST                                                      & 2015 & Cornu          & 1 \\
			\cite{fang2015perfblower}                       & amplification                     &    &    & \X &    &    &    & \X &    &    &    &    & \X &    &    &    &    & Java        & ECOOP                                                    & 2015 & Fang           & 1 \\
			\cite{Zhang2016Isomorphic}                      & augmentation                      &    &    & \X &    & \X &    & \X &    &    & \X &    & \X &    &    & \X &    & Java        & FSE                                                      & 2016 & Zhang          & 3 \\
			\cite{HamletV93}                                & amplification                     &    &    &    & \X &    &    &    &    &    & \X &    &    &    &    &    &    &             & ISSTA                                                    & 1993 & Hamlet         & S \\
			\cite{Xie2006}                                  & augmentation                      &    &    &    & \X &    &    &    &    &    &    & \X &    & \X &    &    &    & Java        & ECOOP                                                    & 2006 & Xie            & 5 \\
			\cite{Joshi07}                                  & amplification                     &    &    &    & \X &    &    &    &    &    & \X &    &    &    & \X &    &    & C           & ESEC/FSE                                                 & 2007 & Joshi          & S \\
			\cite{Mouelhi:2009}                             &                                   &    &    &    & \X &    &    &    &    &    & \X & \X &    &    &    &    &    & Java        & ICST                                                     & 2009 & Mouelhi        & 4 \\
			\cite{Dallmeier2010}                            & enrichment                        &    &    &    & \X &    &    &    &    &    & \X & \X &    &    &    &    &    & Java        & ISSTA                                                    & 2010 & Dallmeier      & 4 \\
			\cite{Carzaniga:2014:COI:2568225.2568287}       & cross-checking                    &    &    &    & \X &    &    &    &    &    & \X &    &    & \X &    &    &    & Java        & ICSE                                                     & 2014 & Carzaniga      & 6 \\
			\cite{xuan2014test}                             & purification                      &    &    &    & \X &    &    &    & \X &    &    & \X &    &    &    &    &    & Java        & FSE                                                      & 2014 & Xuan           & 5 \\
			\cite{xuan:hal-01309004}                        & \stack{purification}{refactoring} &    &    &    & \X &    &    &    &    & \X &    & \X &    &    &    &    &    & Java        & IST                                                      & 2016 & Xuan           & 1\\
			\midrule
		\end{tabular}
	\end{adjustbox}
\end{table*}

Table \ref{tab:summary} shows all the articles considered in this snowballing survey per our inclusion criteria. 
The first column of the table shows the citation information, as given in the ``References'' section. The second column shows the term that the authors use to designate the form of amplification that they investigate. Columns 3 to 18 are divided in three groups. The first group corresponds to the section in which we have included the paper in our survey. The second group corresponds to the different engineering goals that we have identified. The third group captures the different techniques used for amplification in each work. The final columns in the table contain the target programming language, the year and venue in which the paper has been published, the last name of the first author and the iteration of the snowballing process in which the paper was included in the study.

Each row in the table corresponds to a specific contribution.The rows are sorted first by the section in which the papers are included in our study, then by year and then by the last name of the first author. In total, the table contains 49 rows.

One can see that ``augmentation'' (15 contributions), ``generation'' (9 contributions) and ``amplification'' (7 contributions) are the  terms that appear most frequently to describe the approaches reported here. Other similar terms such as ``enrichment'', ``adaptation'' and  ``regeneration'' are used less frequently. 
Most proposals (19 contributions) focus on adding new test cases to the existing test suite. Test amplification in the context of a change or the modification of existing test cases have received comparable attention (16 and 14 contributions respectively). Some techniques that modify existing test cases also target the addition of new test cases (3 contributions) and amplify the test suite with respect to a change (3 contributions). Amplification by runtime modification is the least explored area.

Most works aim at improving the code coverage of the test suite (25 contributions). After that, the main goals are the detection of new faults and the improvement of observability  (13 and 12 contributions respectively). Fault localization, repair improvement and crash reproduction receive less attention (4, 4 and 1 contributions respectively).

47 papers included in the table have been published between 2003 and 2017. One paper was published back in 1993. Between years 2009 and 2016 the number of papers has been stable (mostly four or five per year). In 2014 two extensions to previous works have been published in addition to five original works, making it the year with most publications on the subject.

\begin{figure}
	\centering
	\includegraphics[width=1\linewidth]{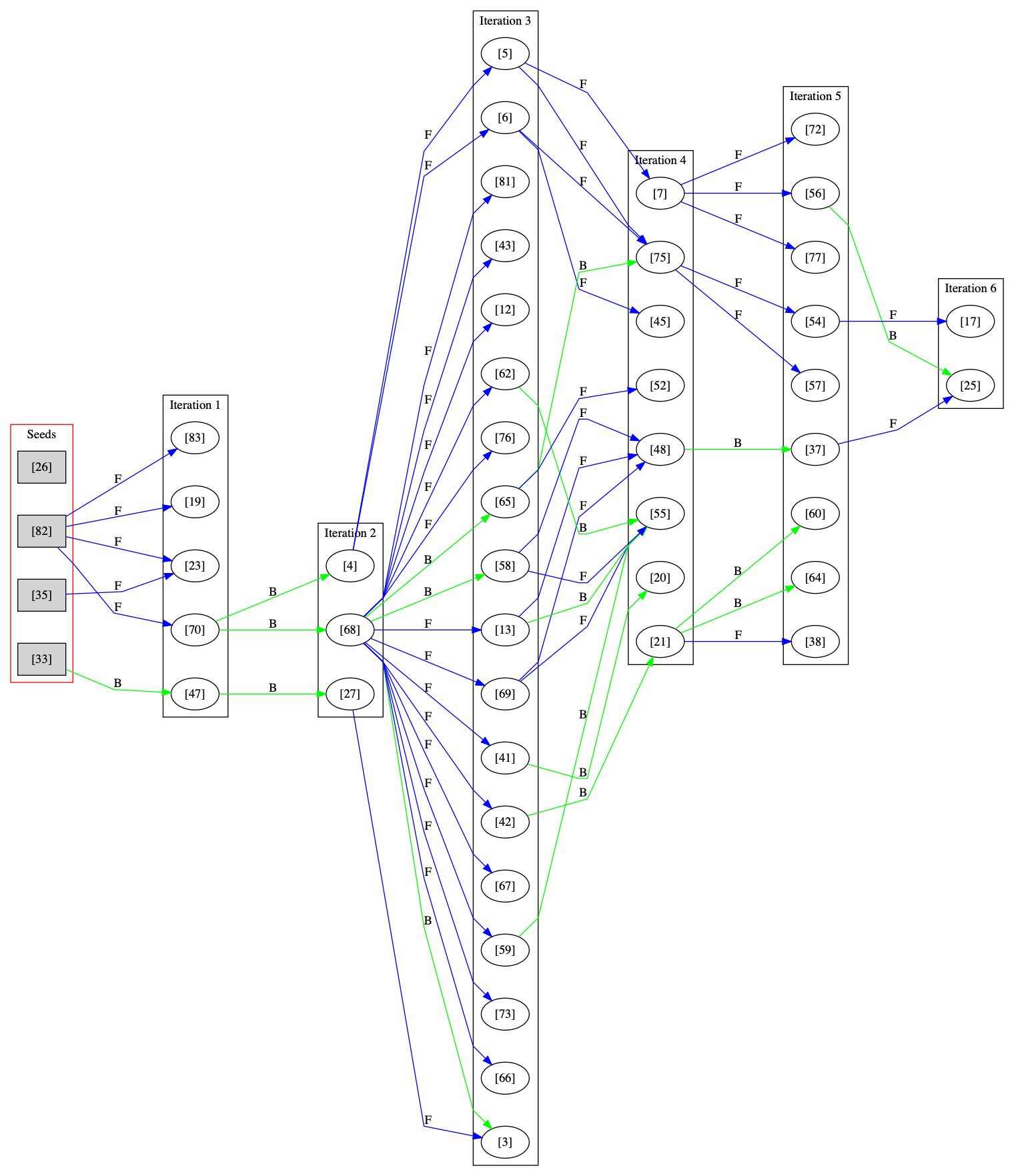}
	\caption{Visualization of the snowballing process. Each node corresponds to a paper included in the study. Seed papers are differentiated form the rest. Papers added in the same iteration are clustered together. \textcolor{blue}{F} blue edges represent forward references. \textcolor{green}{B} represent backward references.}
	\label{fig:citation_graph}
\end{figure}

\rev{Figure \ref{fig:citation_graph} visualizes the snowballing process. Every node of the graph corresponds to a review paper. Seed papers are represented as filled rectangles to distinguish them from the rest. All nodes incorporated in the same iteration are clustered together. The edges shown in the graph correspond to the references we followed for the paper inclusion. Backward references are marked in green and labelled ``B''. For these edges, the origin node cites the target node. Forward references are marked in blue and labelled ``F''. For these edges, the origin node is cited by the target node.}

\subsection{Technical Aspects}

Most works include some form of test or application code analysis (26 and 21 contributions respectively). Notably, the majority of works that add new test cases also include a test code analysis phase. All papers that amplify the test suite with respect to a change also include an application analysis stage. Search-based heuristics and symbolic execution are used to a large extent (12 contributions each), while concolic execution and execution modification are the least used techniques (5 contributions each).

Java programs are the most targeted systems (30 contributions), followed by C programs (12 contributions). JavaScript applications have received very little attention in the area (only one row).

\subsection{Tools for Test Amplification}
\label{sec:open-science}

Most test case amplification papers discussed in this paper are experimental in nature, and are based on a prototype tool. 
For the field to mature, it is good if researchers can reproduce past results, and compare their new techniques against existing ones.
To this extent, we feel that open-science in the form of publicly-available and usable research prototypes is of utmost importance. 

With this in mind, we have surveyed not only the articles, but also the mentioned tools, if any. 
The protocol was as follows. 
First, we looked for a URL in the paper, pointing to a web page containing the code of the tool or experimental data.  
For each URL, one of the authors opened it in a browser between March 1st and March 31st 2018, to check that the page still exists and indeed contains experimental material.

\newcounter{tblerows}
\expandafter\let\csname c@tblerows\endcsname\rownum

\begin{table*}[ht]
	\caption{List of surveyed papers in which we have found a URL related to a tool}
	\label{tab:table:tools:urls}
	\centering
	\small
	\rowcolors{2}{gray!25}{white}
	\begin{tabularx}{\textwidth}{lXX}
		\toprule
		Reference & URL & Observations \\
		\midrule
		\cite{SIR}                                & \url{http://sir.unl.edu}                                          & This is a software repository. It is not a tool for amplification but it is a resource that could be used for amplification.\\
		\cite{Baudry:2006:ITS:1134285.1134299}    & \url{http://www.irisa.fr/triskell/results/Diagnosis/index.htm}    & The URL points only to results. \\
		\cite{bohme2014corebench}                 & \url{http://www.comp.nus.edu.sg/~release/corebench/}              & The website also contains empirical results.\\
		\cite{Carzaniga:2014:COI:2568225.2568287} & \url{http://www.inf.usi.ch/phd/goffi/crosscheckingoracles/}       & \\
		\cite{Dallmeier2010}                      & \url{https://www.st.cs.uni-saarland.de/models/tautoko/index.html} & \\
		\cite{reassert2009}                       & \url{http://mir.cs.illinois.edu/reassert/}                        & \\
		\cite{fang2015perfblower}                 & \url{https://bitbucket.org/fanglu/perfblower-public}              & There is no explicit url in the paper but a sentence saying that the tool is available in Bitbucket. With this information it was easy to find the URL. \\
		\cite{fraser2011evosuite}                 & \url{http://www.evosuite.org/}                                    & Additional materials included. \\
		\cite{marri2010retrofitting}              & \url{https://sites.google.com/site/asergrp/projects/putstudy}     & The website also contains empirical results.\\
		\cite{milani2014}                         & \url{https://github.com/saltlab/Testilizer}                       & \\
		\cite{Pacheco2005}                        & \url{http://groups.csail.mit.edu/pag/eclat/}                      & The website provides basic usage example.\\
		\cite{palikareva2016shadow}               & \url{https://srg.doc.ic.ac.uk/projects/shadow/}                   & The website also contains empirical results.\\
		\cite{pezze2013}                          & \url{http://puremvc.org/}                                         & The paper has been turned into a company. The provided url is the url of this company.\\
		\cite{robetaler2012isolating}             & \url{https://www.st.cs.uni-saarland.de/bugex/}                    & The url lives, but there is no way to download and try the tools. \\
		\cite{xuan:hal-01309004}                  & \url{https://github.com/Spirals-Team/banana-refactoring}          & \\
		\cite{xuanTSE2016Nopol}                   & \url{https://github.com/SpoonLabs/nopol}                          & Still active. \\
		\cite{Zhang2016Isomorphic}                & \url{https://github.com/sei-pku/Ison}                             & \\
		\bottomrule
	\end{tabularx}
\end{table*}

\autoref{tab:table:tools:urls} contains all valid URLs found. Overall, we have identified 17 valid open-science URLs. It may be considered as a low ratio, and we thus call for more open-science and reproducible research in the field of test amplification.

\subsection{Open Questions for Future Research}
\label{subsec:research-question}

Most of the work discussed targets the unit test level, i.e., small tests that verify singles behaviors.
Yet we do not see conceptual barriers to using them in acceptance tests of GUI tests such as Selenium.

\section{Threats to validity}
\label{sec:threats}

Since conducting a survey is a largely manual task, most threats to validity relate to the possibility of researcher bias, and thus to the concern that other researchers might come to different results and conclusions. One general remedy that we adopted to counter this, is to work in a structure way, i.e., by starting from a small set of seed papers, use the citation graph to discover new papers.

In the following, we describe validity threats and discuss the manners in which we attempted to minimize their risk.

\paragraph{Article selection} Test amplification is a relatively new and narrow subject and we found that, as yet, there is no consensus on terminology. Therefore, we used a snowballing survey, which is less likely to be affected by the use of diverse terminologies. This approach is also immune for the issues of keyword based searches, which as has been observed by Brereton et al., can be problematic~\cite{Brereton2007lessons}.

\paragraph{Completeness} We have addressed the threat of selection bias by utilizing the aforementioned snowballing approach. However, some related work could be missing either because they use a very original name for referring to test amplification or because it has not yet been cited by any of the seed papers that we used to start the snowballing effort.

\paragraph{Article categorisation} We have organised the papers in our survey in four categories, through an incremental analysis of the techniques and goals of each paper. The construction of the categorisation is subjective and may be difficult to reproduce. To minimize this risk, two authors performed the categorisation and had to reach consensus, both at the level of creating categories and assigning papers to categories.

\section{Conclusion}
\label{sec:conclusion}

We have studied the literature related to test amplification.
This survey is the first that draws a comprehensive picture of the different engineering goals proposed in the literature for test amplification. In particular, we note that the goal of test amplification goes far beyond maximizing coverage only.
We also give an overview of the different techniques used, which span a wide spectrum, from symbolic execution to random search and execution modification.

We believe that this study will help future PhD students and researchers entering this new field to understand more quickly and more deeply the intuitions, concepts and techniques used for test amplification. Finally, we note the lack of work that tries to compare ``traditional'' test generation (generating test cases from scratch), for which there is a myriad of papers, and test amplification (generating tests from existing tests). We think that sound and systematic experimental comparison of different test creation techniques would be a milestone for the nascent and emerging field of test amplification.

\section*{Acknowledgement}

This work has been partially supported by the EU Project STAMP ICT-16-10 No.731529.

\section*{References}

\bibliographystyle{abbrv}

\bibliography{reference.bib}


\end{document}